\documentclass{article}
\usepackage[]{graphicx}
\usepackage[]{epsfig}

\newcommand{\beq}{\begin{equation}}
\newcommand{\eeq}{\end{equation}}
\newcommand{\beqa}{\begin{eqnarray}}
\newcommand{\eeqa}{\end{eqnarray}}

\begin{document}

\title{Tap Complexity, the Cavity Method and Supersymmetry}
\author{Tommaso Rizzo
\\
\small Center for Statistical Mechanics and Complexity, INFM Roma ``La Sapienza'' 
\\
\small Piazzale Aldo Moro 2, 00185 Roma, 
Italy}

\maketitle
\begin{abstract}
We compute the Bray and Moore (BM) TAP Complexity for the Sherrington-Kirkpatrick model through the cavity method, showing that some essential modifications are needed with respect to the standard formulation of the method. This allows to understand various features recently discovered and to unveil at last the physical meaning of the parameters of the BM theory. We also reconsider the supersymmetric (SUSY) formulation of the problem finding that the BM solution satisfies some proper SUSY Ward identities that are different from the standard ones. 
The SUSY relationships encode the physical meaning of the parameters obtained through the cavity method. The problem of the vanishing prefactor is addressed showing how it can be avoided. 
\end{abstract}
 
\section{Introduction}

The cavity method was introduced nearly twenty years ago in Ref. \cite{MPVc} in order to recover the Parisi solution of the Sherrington-Kirkpatrick (SK) spin-glass model without the replica method and starting explicitly from the hypotheses encoded  in his replica-symmetry breaking (RSB) ansatz \cite{MPV}.
In the last years this method has proven to be a powerful tool to investigate disordered systems with tree-like structure  and optimization problems (see \cite{MP1,MP2,MZ,MZP} and references therein).
In particular it has been possible to obtain a non-perturbative equilibrium solution of the Spin-Glass on the Bethe lattices at the 1RSB level \cite{MP1}. The method has been also used to compute the complexity {\it i.e.} the logarithm of the number of metastable states with free energy higher than the equilibrium one \cite{MP2}. 
It is natural to expect that in the limit of high connectivity this model should become identical to the SK model and the corresponding complexity curve should be equal to the number of stable solutions of the Thouless-Anderson-Palmer (TAP) equations \cite{TAP}.
However the resulting complexity turns out to be different from the TAP complexity computed more than twenty years ago by Bray and Moore (BM) in \cite{BMan} and instead to be equal to another complexity curve yielded by the so-called supersymmetric (SUSY) solution \cite{PP,CGPM,noian,noiquen}.
This has motivated a renewed interest \cite{CGPM,CGPal,noian,noiquen} in the problem of the TAP Complexity starting from the observation that the BM solution violates a supersymmetry (SUSY) \cite{PaSou1,PaSou2,BRST,ZZ} as first noted in \cite{K91}.
Furthermore after the observation \cite{noiquen} that the SUSY solution is unstable in the SK model (except at the lower band edge) both at the annealed and at the quenched level, the BM solutions has remained the unique candidate to describe the TAP complexity and it has becomes mandatory to solve the problems posed by the SUSY violation; as we will briefly report in the following some important results in this direction have been obtained in \cite{ABM} and rigorously confirmed in \cite{PR}. 
In this paper we will address the problem of the formulation of the BM solution within the cavity method obtaining as a byproduct to clarify various issues connected to the SUSY violation and to eventually obtain the physical meaning of the parameters $\lambda$ and $\Delta$ of the theory.

Within the BM computation \cite{BMan} the TAP complexity 
is expressed  as an integral over a certain action that can be evaluated by the saddle point method in the thermodynamic limit. Their solution yields a bell-shaped complexity curve $\Sigma(f)$ describing solutions of the TAP equations with a given free energy.
This action posses a SUSY that is violated by the BM solution \cite{K91}. 
We refer the reader to Ref. \cite{K91,CGPM,noian} for the discussion of the SUSY of the problem and in particular to \cite{PR} whose notation will be adopted in the discussion of section \ref{sec:thesusy}.
This somehow exotic symmetry is related to important physical features of the problem. The first is the Morse theorem that states that the number of solutions of the TAP equations with positive Hessian ({\it e.g.} minima) minus the number of solutions with negative Hessian (the saddles) is a topological invariant equal to one. It is the generalization of the one dimensional result that is easily understood. In particular if we find an exponential number of minima we must find also an exponential number of saddles.
Before the investigations of the last months it was believed that the BM solution described minima of the TAP free energy, therefore an unavoidable question was: were are the saddles?
Furthermore given that those supposed minima of the TAP free energy had a positive definite Hessian it should be possible to continue each one of them upon  continuously changing the external parameters of the TAP equations such as temperature or magnetic field and also upon adding a spin to the system of $N$ spins. The possibility of continuing the solutions poses two problems. The first problem, noted in \cite{noian}, is that starting from the hypothesis that the TAP solutions can be continued one can obtain some relations that are identical to the Ward identities yielded by the SUSY of the problem; thus there is a contradiction between the possibility of continuing the TAP solutions and the fact the BM solution describing them is not SUSY and does not satisfy the corresponding Ward identities.     
The second problem is that since the solutions can be continued upon changing the external parameters they cannot disappear, therefore their number must remain the same at all temperatures and magnetic fields, but this is in contradiction with the fact that the BM complexity varies with temperature and field.
On the other hand if the complexity changes with the temperature and the magnetic field the relevant exponential number of solutions at given values of the external parameters must disappear upon a small change of them and therefore the Hessian of these solutions must have at least a zero eigenvalue. 

These problems of the BM solution have been solved reconsidering the TAP spectrum. In \cite{ABM} it was pointed out that the spectrum contains an isolated eigenvalue besides the continuous positive band and it was checked numerically that this eigenvalue vanishes on the TAP solutions described by the BM solution, this result has been proven rigorously in \cite{PR} and it is precisely a consequence of the SUSY violation.
Following \cite{ABM} and \cite{PR} we recall that  the appearance of the isolated eigenvalue is connected to the fact that the Hessian $X^{-1}$ can be written in the form $A=B+P$ where $P$ is a projector.
Given two symmetric matrices $A$ and $B$ that differ by a projector $P=|\alpha \rangle \langle \alpha |$ we have:
\begin{equation}
A= B + P \longrightarrow \det A= (1+\langle \alpha |B^{-1}|\alpha \rangle  )\det B \label{abp}\ ,
\end{equation}
that is the determinant of the total Hessian $A$ is the product of the determinant of $B$ times a factor depending on $B$ and on the projection vector. The matrix $B$ is essentially the matrix of the interactions and it can be studied through standard random matrix theory \cite{BMsp,Ple1,Ple2,mehta}.  It turns out to have always a positive spectrum with a continuous band of eigenvalues 
whose lower band edge extends down to zero if the following quantity is zero: 
\begin{displaymath}
x_p=1-\beta^2\sum_i(1-m_i^2)^2
\end{displaymath}
The quantity $x_p$ however is different from zero on the BM solution, although it vanishes at the lower band edge in the quenched case where it coincides with the Parisi solution \cite{BMY,noian}.
The inclusion of the projection term modify the eigenvalues of $B$, however the $N-1$ higher eigenvalues remain confined in the original band {\it i.e.} they remain positive,  only the lowest eigenvalue is split out of the band of a finite amount possibly becoming negative. 
Therefore the TAP solutions can only be minima or saddles of order one depending on the sign of the isolated eigenvalue or equivalently of the factor  $(1+\langle \alpha |B^{-1}|\alpha \rangle  )$, if this term is zero the isolated eigenvalue vanishes and the solution is an inflection point. Starting from equation \ref{abp} applied to $B=A-P$ it is easy to prove that the factor controlling the determinant of the Hessian can be also written as:
\begin{equation}
1+\langle \alpha |B^{-1}|\alpha \rangle  ={1 \over 1-\langle \alpha |A^{-1}|\alpha \rangle  }
\label{ABinv}
\end{equation}
Thus we can express the factor controlling the determinant of the Hessian $A$ in term of the susceptibility matrix $X=A^{-1}$, and the key object controlling the sign of the isolated eigenvalue is the quantity $L$ defined as:
\beq
L={1\over N}\sum_{ij}m_iX_{ij}m_j
\eeq
In \cite{PR} it is proven that this quantity is divergent on the BM solution because of the SUSY violation therefore the isolated eigenvalue is zero.
On the numerical ground the situation is not completely clear \cite{Plenum,CGP04} although some evidences in favor of the BM solution and of its predictions have recently exhibited in \cite{CGP04}. Furthermore an argument can be advanced  for the continuity of the TAP Complexity at $T=0$ \cite{argP}, a property that  is satisfied by the BM solution.

The study of the isolated eigenvalue however have not solved all the problems connected to the BM solutions. For instance it is not clear why the whole curve $\Sigma(f)$ is dominated by solutions with a zero eigenvalue and not only the point where it is maximal \cite{PR}.
It is also known \cite{K91} that the prefactor of the exponential contribution to the total complexity vanishes at all orders in power of $1/N$, a result that was extended to the whole curve $\Sigma(f)$ in \cite{PR}. As discussed in \cite{K91} and \cite{PR} this may be a good thing to recover the Morse theorem prediction but leaves open the question of wether we can really identify the 
complexity with the exponential contribution if its prefactor is zero.

Most importantly the presence of a zero eigenvalue complicates the behavior of the TAP solutions upon changing the external parameters, and in particular the possibility or not of continuing them, a property that is crucial to set up the Cavity method in order to apply the theory to a wider class of problems.
In the following we will answer to these questions, as a byproduct  we will eventually obtain the physical meaning  of the parameters  $\lambda$ and $\Delta$ of the BM theory that have remained obscure up to now.

In order to understand the whole features of the BM solution for the TAP complexity we must consider two order parameters, the free energy $f$ and the self-overlap $q$ and not only $f$.
Thus we must study the function $\Sigma(f,q)$ or equivalently the function $\tilde{\Sigma}(u,\lambda_q)$ defined as
\beq
\Sigma(f,q)=\ln \sum_{\alpha} \delta(q-q_{\alpha \alpha}) \delta (f-f_{\alpha})\ \ \ ;\ \ \ 
\Sigma(u,\lambda_q)=\ln \sum_{\alpha}e^{-u f_{\alpha}-\lambda_q q_{\alpha \alpha}}
\eeq
The extremization with respect to $q$ and $f$, in order to compute the total complexity, requires to set $u=0$ and $\lambda_q=0$ in the computation of $\tilde{\Sigma}$, however we will show that {\it we cannot set $\lambda_q=0$ from the beginning but we must study the $\lambda_q \neq 0$ case and  take the limit, indeed $\lambda_q$ appears in products with quantities that are divergent in the limit $\lambda_q\rightarrow 0$ thus yielding a finite contribution. One of these diverging quantities is precisely the parameter $L$ introduced above whose inverse controls the sign of the isolated eigenvalue.} 
On the contrary there are no divergences associated to the parameter $u$ that can be safely set to zero from the beginning. Furthermore the singular behavior associated to the limit $\lambda_q\rightarrow 0$ is present for each $u$ thus explaining why the whole line $\Sigma(f)$ of the BM solution is singular and the prefactor of the exponential is zero.
In the following we will consider only the case $u=0$ and we will study the curve $\Sigma(q)$ or  its Legendre transform $\tilde{\Sigma}(\lambda_q)$. 
We recall that in the thermodynamic limit the two formulations are completely equivalent because the function $\tilde{\Sigma}(\lambda_q)$ defined above is dominated by the solution at a given value of $q$ that fixes a univoque correspondence between $\lambda_q$ and $q$ , the two functions are indeed Legendre transforms one of the other:
\beq
\lambda_q={d \Sigma \over dq}\ ;\ \ \  q=-{d \tilde{\Sigma}\over d \lambda_q}\ ;
\label{legtrasf}
\eeq
\beq
\tilde{\Sigma}=\Sigma -\lambda_q q
\eeq
We will find that the isolated eigenvalue is proportional to $\lambda_q$ thus the determinant of the Hessian of the TAP solution is zero at $\lambda_q=0$ as already shown in \cite{ABM} and \cite{PR} {\it but it is finite for} $\lambda_q\neq 0$. Considering eq. (\ref{legtrasf}) we see that the point $q^{*}$ where the curve $\Sigma(q)$ is maximal separate two regions: on one side we have the minima while on the other side we have the saddles of order one, the two regions touch at $q=q^{*}$ that corresponds to solutions that are inflection points in the TAP landscape.

As soon as $\lambda_q \neq 0$ the corresponding TAP solutions have a non-vanishing determinant of the Hessian thus it is possible to continue them upon changes of the external fields, {\it e.g.} temperature and magnetic field.
In particular we will be able to apply the cavity method studying the continuation of each TAP  solution when a new spin is added.
The use of the cavity method to compute the complexity is discussed in the literature, see {\it e.g.} \cite{MP2} and \cite{noiquen}, the crucial modification presented in the next section is that  to recover the BM prediction for TAP complexity we need to weight the states with the overlap rather than with the free energy. 

In section \ref{sec:thesusy} we will discuss the $\lambda_q\neq 0 $ case within the SUSY context, this will allow to clarify many points, from the behavior of the prefactor to the meaning of the SUSY Ward identities. We will find that the BM solution do satisfy the SUSY Ward identities at each $\lambda_q$, however because of the afore cited divergences in the limit $\lambda_q\rightarrow 0$ these Ward identities are different from those obtained setting simply $\lambda_q=0$ ({\it e.g.} the one considered up to now in the literature see {\it e.g.} \cite{K91,CGPM,noian}), in other words {\it if we carefully take the limit $\lambda_q\rightarrow 0$ we find that the non-SUSY solution  satisfies the correct SUSY identities}.   

Before entering the computation of the TAP complexity within the cavity method we recall what are the modifications that must be made to study the $\lambda_q\neq 0$ case in the standard BM  computation of Ref. \cite{BMan}.
To compute the function $\tilde{\Sigma}(\lambda_q)$ we must simply add to the expression (15) of \cite{BMan} a term equal $\lambda_q q$ and then extremize at fixed $\lambda_q$ with respect to the parameters $q$, $B$, $\Delta$ and $\lambda$ (not to be confused with $\lambda_q$). The resulting equation for $B$, $\Delta$ and $\lambda$ are the same as in the $\lambda_q=0$ case while the equation obtained extremizing with respect to $q$ is
\beq
\lambda_q= -\lambda+\Delta+B-{1\over 2 q}\left( 1- \frac{\langle (\tanh^{-1}m_0-\Delta m_0)^2 \rangle}{q \beta^2}\right)     
\eeq
Consistently upon computing the complexity one finds that $\lambda_q$ and $q$ satisfy the relations (\ref{legtrasf}).

\section{Cavity method for the TAP Complexity}

\subsection{The Continuation of a TAP solution when a new spin is added}

The TAP equations of the system with $N$ spins read:
\beq
-\beta \sum_{i\neq j}m_j +{\beta^2 \over N}\sum_j(1-m_j^2)m_i+\tanh^{-1}m_i =0 
\eeq
We add a new spin $m_0$ to the system and make the following definitions:
\beq
m_i^{(N+1)}=m_i+\delta m_i
\eeq
\beq
\Delta q=Q^{(N+1)}-Q^{(N)}=m_0^2+2 \sum_i m_i \delta m_i +\sum \delta m_i^2
\eeq
\beq
R(\delta m_i)=\tanh^{-1}m_i^{(N+1)}-\tanh^{-1}m_i -{1\over 1-m_i^2}\delta m_i=
{1\over 2}\left({d \over dm_i}{1 \over 1-m_i^2}\right)\delta m_i^2+O\left({1\over N^{3/2}} \right)
\eeq

\beq
X^{-1}_{ij}=\left({1\over \beta} {1\over 1-m_i^2}  +{\beta \over N}\sum_{j=1}(1-m_j^2)\right)\delta_{ij}-J_{ij}-{2 \beta \over N}m_i m_j
\eeq
We also define two fields and three parameters for later use:
\beqa
H_1 & = & \sum_j J_{j0}m_j
\\
H_2 & = & \sum_j J_{j0}(\sum_i X_{ij} m_i)
\\
L & = & {1\over N} \sum_{ij}m_i X_{ij} m_j
\label{defL}
\\
Z_1 & = & {1\over N} \sum_{ij}m_i (X^2)_{ij} m_j
\\
Z_2 & = & {1\over N} \sum_{ij}(X_{ij} m_i)\left({d \over dm_i}{1 \over 1-m_i^2}\right)\sum_s X_{sj}^2
\\
X_{SG} & = & {1\over N} \sum_{j} (X^2)_{jj}
\eeqa
We adopt the convention that all the objects without the label $(N+1)$ are computed on the $N$ system.
From the TAP equations in presence of the spin $m_0$ we can derive an {\it exact} expression for $\delta m_i$ in implicit form:
\beq
\delta m_i=\sum_iX_{ij}\left(J_{j0}m_0-{\beta  \over N}(1-m_0^2)m_j +{\beta \over N}\sum_{s=1}\delta m_s^2 m_j-{\beta \over N}(1-\Delta q)\delta m_j-{1\over \beta}R(\delta m_j)\right)
\label{dmiex}
\eeq
From the previous expression we have:
\beq
\delta m_i=\sum_iX_{ij}J_{j0}m_0+O\left( {1\over N} \right)
\eeq
Therefore
\beq
{1 \over N}\sum_{s=1}\delta m_s^2 ={1\over N}{1 \over N} {\rm Tr} X^2 m_0^2=m_0^2{1\over N}X_{SG}+\dots 
\label{sumdmi2}
\eeq
Using this relation we can easily obtain from the exact expression (\ref{dmiex}) an expression for $\delta m_i$ in explicit form valid to order $1/N$:
\beq
\delta m_i=\sum_jX_{ij}\left(J_{j0}m_0-{\beta  \over N}(1-m_0^2)m_j +m_0^2{\beta \over N}X_{SG}m_j-{m_0^2\over 2 \beta} \left({d \over dm_j}{1 \over 1-m_j^2}\right)\left(\sum_s X_{sj}J_{s0}\right)^2
\right)+O \left( {1 \over N^{3/2}} \right)
\label{dmiap}
\eeq

\subsection{The expression for $m_0$ at order $O(1)$ and the susceptibilities}
The equation for $m_0$ at order $O(1)$ is:
\beq
\tanh^{-1}m_0+\beta^2(1-q)m_0-\beta \sum_i J_{i0}m_i^{(N+1)}=0
\eeq
the field $H_1^{(N+1)}$ can be expressed in term of the field $H_1$:
\beq
H_1^{(N+1)}=H_1+\sum_i{J_{i0}}\delta m_i=H_1+\sum_{ij}{J_{i0}}X_{ij}J_{j0}m_0+O\left({1\over N}\right)=
H_1+m_0{1\over N}{\rm Tr}X+O\left({1\over N}\right)
\eeq
Introducing the variable $B$:
\beq
B=\beta^2 (1-q)-{\beta \over N} {\rm Tr}X \ ,
\label{B}
\eeq
we can express $m_0$ in terms of the field $H_1$ computed on the system with $N$ spins:
\beq
\tanh^{-1}m_0+B m_0-\beta H_1=0\ .
\eeq
In presence of a magnetic field acting on the spin $m_0$ we have
\beq
\tanh^{-1}m_0+B m_0-\beta H_1-\beta h_0=0\ ,
\label{m0h0}
\eeq
and we can obtain the following susceptibilities at order $O(1)$:
\beq
{d m_0\over d h_0}=X_{00}=\beta \left( {1\over 1-m_0^2}+B\right)^{-1}+\dots
\label{X00}
\eeq

\beq
{d^2 m_0 \over d h_0^2}={d \over d m_0}\left( {dh_0 \over dm_0}\right)^{-1}{ d m_0 \over dh_0}=
- \left( {d \over dm_0}{1 \over 1-m_0^2}\right) \beta \left( {1\over 1-m_0^2}+B \right)^{-2}{ d m_0 \over dh_0}+\ldots
\label{sus2}
\eeq
The mixed susceptibility can be obtained from the explicit expression of $\delta m_i$ valid at order $O(1/N)$, equation (\ref{dmiap}):
\beq
X_{i0}^{(N+1)}={d m_i^{(N+1)}\over d h_0}=
\eeq
\beq
=\sum_jX_{ij}\left(J_{j0}X_{00}+{2\beta  \over N}m_0 X_{00}m_j +2m_0 X_{00}{\beta \over N}X_{SG}m_j-{m_0 X_{00}\over  \beta} \left({d \over dm_j}{1 \over 1-m_j^2}\right)\left(\sum_s X_{sj}J_{s0}\right)^2
\right)+O \left( {1 \over N^{3/2}} \right)
\label{mixsus}
\eeq

\subsection{The Self-Overlap shift}
The shift of the self-overlap reads:
\beq
\Delta q=Q^{(N+1)}-Q^{(N)}=m_0^2+2 \sum_i m_i \delta m_i +\sum \delta m_i^2
\eeq
The term $\sum_i m_i \delta m_i$ can be evaluated at order $O(1)$ starting from the expression for $\delta m_i$ at order $O(1/N)$, equation (\ref{dmiap}):
\beq 
\sum_i m_i \delta m_i =
\eeq
\beq
=\sum_{ij} m_i X_{ij}\left(J_{j0}m_0-{\beta  \over N}(1-m_0^2)m_j +m_0^2{\beta \over N}X_{SG}m_j-{m_0^2\over 2 \beta} \left({d \over dm_j}{1 \over 1-m_j^2}\right)\left(\sum_s X_{sj}J_{s0}\right)^2
\right)+O \left( {1 \over N^{1/2}} \right)=
\eeq
\beq
=m_0 H_2-\beta (1-m_0^2)L+m_0^2 \beta X_{SG} L -{m_0^2\over 2 \beta}Z_2+O \left( {1 \over N^{1/2}} \right)
\eeq
The fourth term has been simplified in the following way:
\beq
\sum_{ij} m_i X_{ij}\left({d \over dm_j}{1 \over 1-m_j^2}\right)\left(\sum_s X_{sj}J_{s0}\right)^2={1\over N}\sum_{ij} m_i X_{ij}\left({d \over dm_j}{1 \over 1-m_j^2}\right)\sum_s X_{js}^2+\dots=Z_2+\dots
\eeq
The expression for $\sum_i \delta m_i^2$ was derived above, see equation (\ref{sumdmi2}). Collecting all the term we obtain the following expression for $\Delta q$ valid at order $O(1)$:
\beq
\Delta q=m_0^2+2 m_0 H_2-2 \beta (1-m_0^2)L+2 m_0^2 \beta X_{SG} L -m_0^2{1\over \beta}Z_2+m_0^2X_{SG}+O(1/N^{1/2})
\label{dq}
\eeq

\subsection{The distribution of the Fields}
\label{sec:thefields}
The two fields entering the computation are:
\beqa
H_1 & = & \sum_j J_{j0}m_j
\\
H_2 & = & \sum_j J_{j0}(\sum_i X_{ij} m_i)
\eeqa
Before the reweighting they have a Gaussian distribution with covariances:
\beqa
\langle H_1^2 \rangle_{(N)} & = & {1\over N}\sum_j m_j^2=q
\label{cov1}
\\
\langle H_2^2 \rangle_{(N)} & = & {1\over N}\sum_j(\sum_i X_{ij} m_i)^2={1\over N}\sum_{ij}m_i (X^2)_{ij}m_j=Z_1
\label{cov2}
\\
\langle H_1 H_2\rangle_{(N)} & = & {1\over N}\sum_j m_j(\sum_i X_{ij} m_i)=L
\label{cov3}
\eeqa
After the reweighting the distribution of the fields is proportional to 
\beq
P^{(N+1)}(H_1,H_2)\propto \exp[-{1\over 2}(H_1\ \  H_2)C^{-1}\left(\begin{array}{c}H_1 \\ H_2\end{array}\right)-\lambda_q \Delta q]
\label{distf0}
\eeq
where according to eq. (\ref{cov1}),(\ref{cov2}) and (\ref{cov3}):
\beq
C=\left( \begin{array}{cc} q & L \\ L & Z_1\end{array}\right)
\label{C}
\eeq
At the end of the computation we expect to find that $\lambda_q$ is proportional to the derivative of the complexity with respect to $q$.
\beq
\lambda_q={d \Sigma\over dq}
\eeq
The variation of the self-overlap $\Delta q$ was computed above, see equation (\ref{dq}). Neglecting the irrelevant constant term it can be written as:
\beq
\Delta q={a \over 2}m_0^2+2 m_0 H_2 \ \ \ {\rm with} \ \ \  {a \over 2}=1+2 \beta L+2 \beta X_{SG}L+X_{SG}-{1\over \beta} Z_2
\eeq
The magnetization is univoquely determined by $H_1$ according to the previously derived relation:
\beq
\tanh^{-1}m_0+B m_0-\beta H_1=0
\eeq
Performing the integration over $H_2$ and changing variable from $H_1$ to $m_0$ we obtain that the reweighted distribution of $H_1$ is proportional to:
\beq
\left({1\over 1-m_0^2}+B\right)\exp\left[ -{ (\tanh^{-1}m_0+B m_0+ 2 \beta L \lambda_q m_0)^2 \over 2 q \beta^2}-\lambda_q {a\over 2}m_0^2+2 Z_1 \lambda_q^2 m_0^2\right]dm_0
\eeq
We introduce the new variables:
\beq
\Delta= - B-2 \beta \lambda_q L
\label{delta}
\eeq
\beq
\lambda=-\lambda_q {a\over 2}+2 Z_1 \lambda_q^2 
\label{lambda}
\eeq
And the reweighted distribution takes the BM form:
\beq
P^{(N+1)}(m_0)=K \left({1\over 1-m_0^2}+B\right)\exp\left[ -{ (\tanh^{-1}m_0- \Delta m_0)^2 \over 2 q \beta^2}+\lambda m_0^2\right]dm_0
\label{distf}
\eeq
Where K is the normalization constant.

A word of caution is in order here. The fluctuations of the field $H_1$ and $H_2$ considered in equations (\ref{cov1}-\ref{cov3}) and described by the distribution function (\ref{distf0}) are not fluctuations over the disorder but over the different TAP solutions at a given realization of the disorder. In general  this leads to the fact that the fields before the reweighting are Gaussian but with  non-zero means $H_{1J}$ and $H_{2J}$ depending on the disorder, see eq. V.25 at pg. 71 in \cite{MPV}. 
Then the average over the disorder is carried on  considering the distribution function of $H_{1J}$ and $H_{2J}$ that are again Gaussian with zero-mean but with a covariance matrix that is in general different from zero. In the standard case, see eq. V.37 in \cite{MPV}, the variance is proportional to the overlap between different states and analogous relationship can be obtained in our case. However, if we assume that different TAP solutions of the same sample are uncorrelated, the disorder variances of $H_{1J}$ and $H_{2J}$ becomes zero, that is we have precisely $H_{1J}=0$ and $H_{2J}=0$ on each sample. This corresponds in the replica language to make a 1RSB ansatz with $q_0=0$. If $H_{1J}$ and $H_{2J}$ are zero on each sample, equation (\ref{distf0}) can be identified with the distribution over the TAP solutions and over the disorder. Indeed in the following we make this identification. 
This is correct as soon as we want to reproduce the result of the annealed computation of the TAP complexity. Instead if we keep non-zero $H_{1J}$ and $H_{2J}$ we can reproduce the so-called replica-symmetric quenched computation of the complexity of BM, see equation (19,20,21) in \cite{BMan}. We recall however that as shown in \cite{BMan} the annealed computation is correct for what concerns the total complexity and one can hope that this simplification holds in other models as well.

\subsection{The Self-Consistency Equations}
The self-consistency equation for the various parameters of the theory are:
\beqa
q & = & \langle m_0^2\rangle
\\
L & = & \langle m_0 \sum_i X_{i0}^{(N+1)}m_i^{(N+1)}  \rangle
\\
Z_1 & = & \langle ( \sum_i X_{i0}^{(N+1)}m_i^{(N+1)})^2   \rangle
\\
Z_2 & = & \left\langle \sum_i X_{i0}^{(N+1)}m_i^{(N+1)} \left( {d \over dm_0}{1\over 1-m_0^2}\right) \left(\sum_{j=0}(X_{j0}^{(N+1)}\right)^2  \right\rangle
\\
{1\over N}{\rm Tr}X& = & \langle X_{00}\rangle
\\
X_{SG}& = & \langle X_{00}^2  \rangle+\langle \sum_{j=1}(X_{j0}^{(N+1)})^2 \rangle
\eeqa
The first equation is simply the equation for $q$ in the BM theory.
The self-consistency equation for ${\rm Tr}X/N$ gives the BM equation for $B$, indeed recalling the definition of $B$, equation (\ref{B}), and the expression of $X_{00}$, equation (\ref{X00}), we obtain
\begin{equation}
B=\beta^2(1-q)-\beta \langle X_{00} \rangle=\beta^2 \left(1-q-\langle \left( {1\over 1-m_0^2}+B  \right)^{-1}\rangle\right)
\end{equation}
Recalling the expression for $X_{0i}^{(N+1)}$, equation (\ref{mixsus}) we can evaluate the equation for $X_{SG}$:
\beq
X_{SG} =  \langle X_{00}^2  \rangle+\langle \sum_{j=1}(X_{j0}^{(N+1)})^2 \rangle
\eeq
\beqa
X_{SG} & = &   \langle \beta^2\left( {1\over 1-m_0^2}+B  \right)^{-2 }\rangle+\langle \beta^2\left( {1\over 1-m_0^2}+B  \right)^{-2}\sum_{j=1}\left(\sum_i J_{i0}X_{ji}+O(1/N)\right)^2 \rangle=
\nonumber
\\
 & = & \langle \beta^2\left( {1\over 1-m_0^2}+B  \right)^{-2 }\rangle(1+X_{SG})
\label{caxg}
\eeqa
Therefore:
\beq
X_{SG}=\frac{\langle \beta^2 \left( {1\over 1-m_0^2}+B  \right)^{-2 }\rangle}{1-\langle \beta^2 \left( {1\over 1-m_0^2}+B  \right)^{-2 }\rangle}
\label{XSG}
\eeq
In the case $B=0$ the expression simplifies to:
\beq
X_{SG}={1\over x_p}-1
\eeq
with
\beq
x_p=1-\langle \beta^2 (1-m_0^2)^2\rangle
\eeq
As in \cite{MPV} we can compute the object:
\begin{equation}
X_{SG}-\langle X_{00}^2  \rangle={(1-x_p)^2 \over x_p}
\end{equation}
In this way we recover the stability condition $x_p \geq 0$ of the BM theory in the $B=0$ case as a positivity condition over $X_{SG}-\langle X_{00}^2  \rangle$.

In order to compute the self-consistency equations for the parameter $L$, $Z_1$ and $Z_2$ we need to determine the quantity $\sum_i X_{i0}^{(N+1)}m_i^{(N+1)}$ at order $O(1)$. This can be done starting from the expression of $\delta m_1$ and $X_{0i}^{(N+1)}$ derived above, that is equation (\ref{dmiap}) and (\ref{mixsus}). We don't report the result that turns out to be equal to half the derivative of $\Delta q$ with respect to $h_0$ as it should since $X_{i0}^{(N+1)}=dm_i^{(N+1)}/dh_0$. To proceed in the computation it is also useful to notice that according to equation (\ref{m0h0}) derivatives with respect to $h_0$ are equivalent to derivatives with respect to $H_1$ therefore we have:
\beq 
\sum_i X_{i0}^{(N+1)}m_i^{(N+1)}={1\over 2}{d \Delta q\over d h_0}=
{1\over 2}{d \Delta q\over d H_1}
\label{xm}
\eeq
With this equation we can perform integration by parts and do not need the explicit expression  of $\sum_i X_{i0}^{(N+1)}m_i^{(N+1)}$. We rewrite the selfconsistency equations for $L$, $Z_1$ and $Z_2$ as:

\beqa
L & = & {1\over 2}\langle m_0 {d \Delta q\over d H_1}\rangle
\\
Z_1 & = & {1\over 4}\left\langle \left({d \Delta q\over d H_1}\right)^2   \right\rangle
\\
Z_2 & = & \left\langle {1\over 2}{d \Delta q\over d h_0}\left( {d \over dm_0}{1\over 1-m_0^2}\right)\beta^2 \left( {1\over 1-m_0^2}+B  \right)^{-2 }\right\rangle(1+X_{SG})
\label{Z2}
\eeqa
In the last expression we have used the expression for $\sum_{i=0}(X_{0i}^{(N+1)})^2$ previously derived in the computation of $X_{SG}$ see equation (\ref{caxg}).

\subsection{The $L$ Equation}
The Self-consistency equation for the parameter $L$ reads
\beq
L  =  {1\over 2}\langle m_0 {d \Delta q\over d H_1}\rangle
\eeq
To compute this expression we will use integration by parts. In the following we will use the shorthand expression for the Gaussian part of the distribution of the fields:
\beq
G[H_1,H_2]=\exp[-{1\over 2}(H_1\ \  H_2)C^{-1}\left(\begin{array}{c}H_1 \\ H_2\end{array}\right)]
\eeq
where $C$ is defined in equation (\ref{C}).
This is essentially the distribution of $H_1$ and $H_2$
  before  the reweighting. We also recall that $K$ is the normalization constant of the reweighted distribution.
Integrating by parts we have:
\beqa
L & = & -K \int m_0 {1\over 2 \lambda_q}G[H_1,H_2]{d \over d H_1}\exp[-\lambda_q \Delta q ]dH_1dH_2=
\\
 & = & {1\over 2 \lambda_q}\left( \langle X_{00} \rangle + K \int m_0\left( {d \over d H_1}G[H_1,H_2]\right)\exp[-\lambda_q \Delta q ]dH_1dH_2
 \right)
\eeqa
Taking the derivative of $ dG[H_1,H_2]/dH_1$ and making the integration over $H_2$ we obtain
\beq
2 \beta \lambda_q L= \beta^2 (1-q)-B-\langle m_0 \frac{\tanh^{-1}m_0+B m_0+ 2 \beta L \lambda_q m_0}{q} \rangle
\eeq
Recalling the change of variables 
\beq
\Delta=-B-2 \beta \lambda_q L
\eeq
we obtain
\beq
\Delta=-{\beta^2 \over 2}(1-q)+{1\over 2 q}\langle m_0 \tanh^{-1}m_0 \rangle
\eeq
Thus the selfconsistency equation for $L$ gives an equation identical to the one obtained in the BM theory extremizing with respect to $\Delta$, see \cite{BMan}.

\subsection{The $Z_1$ equation}

The Self-consistency equation for the parameter $Z_1$ reads
\beq
Z_1  =  {1\over 4}\left\langle \left({d \Delta q\over d H_1}\right)^2   \right\rangle
\eeq
Integrating by parts we have:
\beqa
Z &=& {K\over 4} \int \left({d \Delta q\over d H_1}\right)^2   G[H_1,H_2]\exp[-\lambda_q \Delta q]dH_1dH_2=
\nonumber
\\
& = & {K \over 4 \lambda_q^2} \int \left({d^2 \over d H_1^2}G[H_1,H_2]\right)\exp[-\lambda_q \Delta q]dH_1dH_2+
\nonumber
\\
& + & {K \over 4 \lambda_q} \int{d^2 \Delta q\over d H_1^2}   G[H_1,H_2]\exp[-\lambda_q \Delta q]dH_1dH_2
\label{Z}
\eeqa
A simple computation shows that the first term gives the following contribution:
\beq
{K \over 4 \lambda_q^2} \int \left({d^2 \over d H_1^2}G[H_1,H_2]\right)\exp[-\lambda_q \Delta q]dH_1dH_2={1 \over 4 \lambda_q^2}\left( -{1\over q}\left( 1- \frac{\langle (\tanh^{-1}m_0-\Delta m_0)^2 \rangle}{q \beta^2}\right)      \right) 
\eeq
The second term in (\ref{Z}) requires more care. We start recalling the expression for $\Delta q$, equation (\ref{dq}), neglecting as usual the constant term. As in subsection \ref{sec:thefields} we write it as
\beq
\Delta q={a \over 2}m_0^2+2 m_0 H_2 \ \ \ {\rm with} \ \ \  {a \over 2}=1+2 \beta L+2 \beta X_{SG}L+X_{SG}-{1\over \beta} Z_2
\label{aa}
\eeq
Instead of considering the derivatives with respect to $H_1$ we consider the derivatives with respect to $h_0$ as originally done in equation (\ref{xm}):
\beq
{d^2 \Delta q\over d H_1^2}=
{d^2 \Delta q\over d h_0^2}={d \over dh_0}\left( {d \Delta q\over d m_0}{d m_0\over d h_0}  \right)={d \over d h_0} \left( (a m_0+2 H_2){dm_0\over dh_0}\right)=a \left({dm_0\over dh_0}\right)^2+(2 m_0+ 2 H_2){d^2 m_0\over dh_0^2}
\label{dq2}
\eeq
The second term can be simplified using equation (\ref{sus2}):
\beqa
(2 m_0+ 2 H_2){d^2 m_0\over dh_0^2}& = &
- \left( {d \over dm_0}{1 \over 1-m_0^2}\right) \beta \left( {1\over 1-m_0^2}+B \right)^{-2}(2 m_0+ 2 H_2){ d m_0 \over dh_0}=
\\
& = & - \left( {d \over dm_0}{1 \over 1-m_0^2}\right) \beta \left( {1\over 1-m_0^2}+B \right)^{-2}{d \Delta q\over d h_0}
\eeqa
Recalling the self-consistency equation for $Z_2$, equation (\ref{Z2}), we have that the average of the second term in (\ref{dq2}) is simply given by:
\beq
\langle (2 m_0+ 2 H_2){d^2 m_0\over dh_0^2} \rangle = -{2 Z_2\over \beta}{1\over 1+X_{SG}}
\eeq
To evaluate the average of the first term in (\ref{dq2}) we must recall the expression of the susceptibility, equation (\ref{X00}), and the expression of $X_{SG}$, equation (\ref{XSG}), we have 
\beq
\left\langle \left({dm_0\over dh_0}\right)^2 \right\rangle = {X_{SG} \over 1+X_{SG}}
\eeq
Summing up the two terms and recalling the expression for $a$, equation (\ref{aa}), we obtain:
\beq
\left\langle {d^2 \Delta q\over d h_0^2} \right\rangle=a {X_{SG} \over 1+X_{SG}}-{2 Z_2\over \beta}{1\over 1+X_{SG}}=a-2-4 \beta L
\eeq
Thus equation (\ref{Z}) reads
\beq
Z_1={1 \over 4 \lambda_q^2}\left( -{1\over q}\left( 1- \frac{\langle (\tanh^{-1}m_0-\Delta m_0)^2 \rangle}{q \beta^2}\right)      \right) 
+{1\over 4 \lambda_q}(a-2-4 \beta L)
\eeq
Multiplying both sides by $2 \lambda_q^2$ we obtain:
\beq
2 \lambda_q^2 Z_1= -{1\over 2 q}\left( 1- \frac{\langle (\tanh^{-1}m_0-\Delta m_0)^2 \rangle}{q \beta^2}\right)     
+\lambda_q({ a\over 2}-1-2 \beta L)
\eeq
Now recalling the definition of the variables $\lambda$ and $\Delta$ introduced in subsection \ref{sec:thefields}:
\beq
\Delta= - B-2 \beta \lambda_q L
\eeq
\beq
\lambda=-\lambda_q {a\over 2}+2 Z_1 \lambda_q^2 
\eeq
We obtain the equation:
\beq
\lambda_q= -\lambda+\Delta+B-{1\over 2 q}\left( 1- \frac{\langle (\tanh^{-1}m_0-\Delta m_0)^2 \rangle}{q \beta^2}\right)     
\eeq
This is precisely the equation corresponding to the extremization of the BM action  at a fixed value of the selfoverlap $q$ reported in the introduction and we have the correct identification between $\lambda_q$ and the derivative of $\Sigma(q)$ with respect to $q$:
\beq
\lambda_q={d \Sigma \over dq}
\eeq
We do not report the explicit expression of the complexity that can be obtained taking into account the rescaling of the temperature associated to the process of adding a spin without changing the couplings \cite{MPV}. 

\section{On Supersymmetry}
\label{sec:thesusy}
By constraining the solutions to have a given overlap $q$ or equivalently by weighting the sum over solutions with a weight proportional to $e^{-\lambda_q q}$ we break the supersymmetry of the problem. Considering the bell shaped curve described by $\Sigma(q)$ we see that at the maximum the action is SUSY while for $q$ different from the maximum value $q^*$ or equivalently for $\lambda_q\neq 0$ the action is not SUSY but the SUSY violation takes a very simple form. 

In the following we will see that at any value of $\lambda_q$ it is possible to write some SUSY Ward Identities that reduce to the SUSY relations in the limit $\lambda_q\rightarrow 0$. In this relations the parameter $\lambda_q$ multiplies some quantities that diverge in the limit $\lambda_q\rightarrow 0$ therefore taking carefully this limit we obtain relations that are different from those obtained simply putting $\lambda_q=0$ in the equations. 

The point $\Sigma(q^*)$ corresponding to the maximum of $\Sigma (q)$ is singular, there is a zero eigenvalue that is definitively different from zero a finite $\lambda_q$ and changes sign at $\lambda_q=0$. This can be clearly seen because this eigenvalue is proportional to the inverse of the physical parameter $L$ that diverges as $\lambda_q \rightarrow 0$ since $L \propto\Delta/\lambda_q$. Therefore for each $\lambda_q\neq 0$ the determinant of the TAP solutions is strictly different from zero and we can apply the cavity method considering the continuation of each solution when a new spin is added. 
Notice that the sign of the determinant of the solutions is univoquely defined on the left and on the right of $q^*$ therefore we can safely remove the modulus of the determinant from the standard computation of the complexity and the integral of the macroscopic action will also change sign at $q^*$.  

At finite $N$ the only points where solutions can rise or die are on the line $\lambda_q=0$, then adding or removing spins the solutions move on the curve $\Sigma(q,f)$ since their self-overlap $q$ and free energy $f$ change with $N$ thus leading to the correct behavior of the solutions as $e^{N\Sigma(q,f)}$.

Since the SUSY is broken for $\lambda_q\neq 0$ the determinant of the fermionic fluctuation is no longer zero. Thus at each $\lambda_q\neq 0$ we have a finite prefactor to the exponential contribution, whose expansion in power of $1/N$ vanishes at all orders in the limit $\lambda_q \rightarrow 0$. 
If one does not consider the $\lambda_q\rightarrow 0$ limit the vanishing of the prefactor at $\lambda_q=0$ is a problem since it is not clear if this change the behavior of the complexity. One should be able to prove that even if its expansion vanishes at all orders the prefactor is definitively different from zero at each finite $N$. Notice that this can happen for instance if the prefactor takes the form $e^{-a N}$ thus changing the exponential contribution.
Instead the fact that the prefactor is always finite at $\lambda_q\neq 0$ 
means that the exponential contribution is dominant and allows us to bypass this problem, indeed the identification of the total complexity with the BM  complexity can be safely obtained considering the limit $\lambda_q\rightarrow 0$ of $\tilde{\Sigma}(\lambda_q)$ or equivalently the limit $q\rightarrow q^*$ of $\Sigma(q)$.

At the macroscopic level the SUSY-like relationships we will obtain are not relationships between the macroscopic parameters, instead they are relationships that connect the macroscopic bosonic order parameters {\it e.g.} $\lambda$ and $\Delta$ with averages of the fermionic macroscopic variables. In this way they yield independently of the cavity method the physical meaning of  $\Delta$ and $\lambda$ obtained in the previous section. Indeed the parameter $L$, $Z_1$ and $Z_2$ requires averages over the fermionic variables to be computed and the BRST relationship we will derive are equivalent to the definitions (\ref{delta}) and (\ref{lambda}).  

\subsection{Microscopic Supersymmetry}

If we consider the sum over solutions of the TAP equations weighting each solutions with a weigh proportional to $e^{-\lambda_q q}$ we obtain that the result can be expressed (removing the modulus of the determinant) as an integral over the following action:
\beq
S=S(\{\overline{\psi_i},\psi_i,m_i,x_i\})-\lambda_q \sum_i m_i^2
\eeq
Here and in the following section we will adopt the notation of \cite{PR}.
The first term is invariant under the microscopic BRST derivative $D_{\rm micro}$ induced by the following transformation \cite{BRST,PaSou1,PaSou2}:
\beq
\delta m_i = \epsilon\, \psi_i \quad\quad
\delta \bar\psi_i = -\epsilon\, x_i \quad\quad
\label{brst}
\eeq
\beq
D=\, \psi_i \partial_{m_i}-\, x_i \partial_{\bar\psi_i}
\label{brst2}
\eeq
The action is no longer invariant but the variation is easily computed from the previous relations:
\beq
DS=-2 \lambda_q \sum_i m_i \psi_i
\eeq
As a consequence the average of a BRST derivative is no longer zero but it is given by:
\beq
\langle DO\rangle=\langle  DS O \rangle=-2 \lambda_q \langle \sum_i m_i \psi_i
 O \rangle
\label{DO}
\eeq
Thus we can obtain Ward identities also at $\lambda_q\neq 0$, if we put $O=m_i\overline{\psi_j}$ we obtain the following relationship:
\beq
-\langle m_i x_j \rangle=\langle \overline{\psi_j} \psi_i\rangle-2 \lambda_q \langle m_i \sum_k \overline{\psi_j}\psi_k m_k\rangle
\label{b1}
\eeq
If we set $\lambda_q=0$ in the previous relation we obtain a standard SUSY relationship, instead it is crucial to consider the limit $\lambda_q\rightarrow 0$ because the term multiplied by $\lambda_q$ diverges in this limit leading to a finite result.

Let us show that the previous relation follows naturally if we assume that the solutions can be continued. In this case we have that:
\beqa
{\partial \over \partial h_j}\left(\sum_{\alpha}m_i^{\alpha}e^{-\lambda_q q_{\alpha \alpha}}\right) & = & \sum_{\alpha}{\partial m_i^{\alpha} \over \partial h_j}e^{-\lambda_q q_{\alpha \alpha}}-\sum_{\alpha}m_i^{\alpha}\lambda_q {\partial q_{\alpha \alpha} \over \partial h_j}e^{-\lambda_q q_{\alpha \alpha}}=
\nonumber
\\
& = & \sum_{\alpha }{\partial m_i^{\alpha} \over \partial h_j}e^{-\lambda_q q_{\alpha \alpha}}-2\lambda_q\sum_{\alpha k}m_i^{\alpha} X^{\alpha}_{jk}m_k^{\alpha}e^{-\lambda_q q_{\alpha \alpha}}
\label{b2}
\eeqa
Recalling the following relationships we can easily see the equivalence between equation (\ref{b1}) (obtained considering the BRST relationship) and equation (\ref{b2}) (obtained assuming the possibility of continuing the TAP solutions):
\beq
{\partial \langle m_i \rangle \over \partial h_j}=-\langle m_i x_j\rangle
\eeq
\beq
\langle \overline{\psi_j} \psi_i\rangle=\sum_{\alpha }{\partial m_i^{\alpha} \over \partial h_j}e^{-\lambda_q q_{\alpha \alpha}}
\eeq
\beq
-2 \lambda_q \langle m_i \sum_k \overline{\psi_j}\psi_k m_k\rangle=-2\lambda_q\sum_{\alpha k}m_i^{\alpha} X^{\alpha}_{jk}m_k^{\alpha}e^{-\lambda_q q_{\alpha \alpha}}
\label{Lmp}
\eeq
If we put $O=x_i \overline{\psi_j}$ in (\ref{DO}) we obtain the generalization of another well-known BRST relation to the case $\lambda_q\neq 0$:
\beq
\langle x_i x_j\rangle=2 \lambda_q \langle x_i \sum_k \overline{\psi_j}\psi_k m_k \rangle
\label{c1}
\eeq
On the other hand if we assume that the TAP solutions relevant for the weighted average $\sum_\alpha \exp[-\lambda_q q_{\alpha \alpha}]$ can be univoquely continued we have that:
\beq 
{\partial^2 \over \partial h_i \partial h_j}\sum_{\alpha}e^{-\lambda_q q_{\alpha \alpha}}={\partial \over \partial h_i}\sum_{\alpha}\left(-\lambda_q {\partial q_{\alpha \alpha}\over \partial h_j} e^{-\lambda_q q_{\alpha \alpha}}\right)={\partial \over \partial h_i}\sum_{\alpha}\left(-2\lambda_q \sum_k X_{jk}^{\alpha}m_k^{\alpha}e^{-\lambda_q q_{\alpha \alpha}}\right)
\label{c2}
\eeq
The equivalence between equations (\ref{c1}) and (\ref{c2}) can be proven considering the following equations:
\beq 
{\partial^2 \over \partial h_i \partial h_j}\sum_{\alpha}e^{-\lambda_q q_{\alpha \alpha}}=\langle x_i x_j\rangle
\eeq
\beq
{\partial \over \partial h_i}\sum_{\alpha}\left(-2\lambda_q \sum_k X_{jk}^{\alpha}m_k^{\alpha}e^{-\lambda_q q_{\alpha \alpha}}\right)=-2 \lambda_q {\partial \over \partial h_i}\langle \sum_k \overline{\psi_j}\psi_k m_k \rangle=2 \lambda_q \langle x_i \sum_k \overline{\psi_j}\psi_k m_k \rangle
\eeq
\subsection{The SUSY relationships encode the physical meaning of the parameters}
By summing equation (\ref{b1}) over $i=j$ we obtain
\beq
-\sum_i\langle m_i x_i \rangle-\sum_i\langle \overline{\psi_i} \psi_i\rangle=-2 \lambda_q \langle \sum_{ik}m_i \overline{\psi_i}\psi_k m_k\rangle
\label{summx}
\eeq
Recalling equation (\ref{Lmp}) we see that the r.h.s. it is nothing but the parameter $-2 \lambda_q L$ where $L$ was defined in (\ref{defL}). On the other hand the l.h.s. of the previous equation can be expressed in term of the macroscopic parameters $\Delta$, $B$ and $\lambda$, we have that it is equal to 
\beq
-\sum_i\langle m_i x_i \rangle-\sum_i\langle \overline{\psi_i} \psi_i\rangle=\langle{\Delta +B \over\beta}\rangle
\eeq
Thus equation (\ref{summx}) is equivalent to 
\beq
\langle \Delta + B \rangle= -2 \beta \lambda_q \langle L \rangle
\label{find}
\eeq
Therefore we see that the BRST relationships allow to recover the physical meaning on the parameters $\Delta$, $B$ and $\lambda$ of the standard TAP complexity computation without the cavity method. They relate these parameters to the physical parameters $L$, $Z_1$ and $Z_2$ and the relation is precisely the one that is obtained within the cavity method indeed the last equation is equivalent to equation (\ref{delta}) that is the definition of $\Delta$ within the cavity method. 
Notice once again the importance of taking the limit $\lambda_q\rightarrow 0$ instead of simply setting $\lambda_q=0$. Since $L$ is divergent in this limit the r.h.s. of (\ref{find}) remains finite instead if we put it to zero we obtain the standard SUSY solution $\Delta+B=0$.
In a similar way we can use the SUSY relation (\ref{c1}) to uncover the physical meaning of the parameter $\lambda$ of the standard complexity computation in terms of the physical parameters $L$, $Z_1$, $Z_2$ and $X_{SG}$, obtaining a relation equivalent to equation (\ref{lambda}) obtained within the cavity method. 

\subsection{Macroscopic Supersymmetry}

The results of the previous sections can be recovered also considering the macroscopic action obtained within the computation of the complexity. This depends on four bosonic and four fermionic parameters $\{r,t,\lambda,q,\overline{\rho},\rho,\overline{\mu},\mu\}$. In presence of a forcing term $\lambda_q q$ the action is
\beq
S=S_0-\lambda_q q
\label{totmac}
\eeq
The first term is invariant under a macroscopic BRST transformation given by:
\begin{equation}
\left\{
\begin{array}{ccc}
\delta \mu & = & {2
 \lambda \over \beta} \epsilon
\\
\delta \rho & = & \left( {r  \over \beta }-{t \over \beta} \right)\epsilon
\\
\delta \overline{\mu} & = & 0
\\
\delta \overline{\rho} & = & 0 
\\
\delta q & = & -\epsilon
 {2 \overline{\mu} \over \beta}
\\
\delta \lambda  & = & 0
\\
\delta r & = & -\epsilon (2 \beta \overline{\mu}-\beta \overline{\rho})
\\
\delta t & = & \epsilon (-2 \beta \overline{ \mu }-\beta \overline{\rho})
\end{array}
\right. 
\label{macro}
\end{equation}
Therefore the variation of the total action (\ref{totmac}) is simply given by: 
\beq
\epsilon DS=-\lambda_q \delta q=\epsilon {2 \lambda_q  \over \beta}\overline{\mu} 
\eeq
Much as in the previous sections we have that the averages of a BRST derivative
satisfies the following relation
\beq
\langle DO\rangle=\langle DS O \rangle=-{2 \lambda_q \over \beta} \langle 
 \overline{\mu} O \rangle
\eeq
Substituting for $O$ and looking at the first two relations in (\ref{macro}) we have that 
\beqa
O=\mu & \longrightarrow & \langle {2 \lambda \over  \beta} \rangle= -{2 \lambda_q \over  \beta} \langle \overline{\mu}\mu \rangle
\\
O=\rho & \longrightarrow & \langle {r-t \over  \beta} \rangle= -{2 \lambda_q \over  \beta} \langle \overline{\mu}\rho \rangle
\eeqa
As was derived in \cite{PR} the r.h.s. of the previous equation is proportional to  the physical parameter $L$ while skipping from the variable $\{r,t,\lambda,q\}$ to the BM variables $\{\Delta,B,\lambda,q\}$ the r.h.s.  is simply $(\Delta+B)/\beta$, therefore, much as in the previous subsection, we obtain that the BRST relations within the macroscopic formalism encode the physical meaning of the parameters $\Delta$ and $\lambda$ obtained within the cavity method.

\section{Discussion}
Within the cavity method we have been able to eventually obtain the physical interpretation of  the parameters $\Delta$ and $\lambda$ of the BM theory in terms of physical objects like $L$, $Z_1$, $Z_2$ and $X_{SG}$. 
The extremization with respect to $q$ corresponds to the case $\lambda_q=0$. However in this limit $\Delta$ and $\lambda$ remain finite because  the corresponding combinations of physical parameters are diverging.
In particular we have 
\beq
L =  {1\over N} \sum_{ij}m_i X_{ij} m_j \propto {\Delta \over \lambda_q}
\eeq
Therefore $L$ diverges at $\lambda_q=0$.
As we recalled in the introduction $L$ it is responsible for the behavior of the isolated eigenvalue and its divergence causes this eigenvalue to be zero.
Thus in the space of the parameters $(u,\lambda_q)$, conjugated respectively to the free energy and to the overlap, the extremization corresponds to set $u=0$ and $\lambda_q=0$, but 
the line $\lambda_q=0$ is a singular line, and we must carefully take the limit. Instead the line $u=0$ is regular and we can set $u=0$ from the beginning as we did here.
It would be interesting to check if the presence of two fields $H_1$ and $H_2$, instead of simply one, and the special role played by the self-overlap are connected to the fact that the equivalent replica theory requires the two-group ansatz \cite{BMan,BM2g,PP,noian}. 
The Formulation of the theory within the cavity method opens the way to applications to different models \cite{MP1,MP2} and optimization problems \cite{MZ,MZP}. The case of the Bethe lattice is currently under investigation.

Within the SUSY framework we have shown that setting $\lambda_q\neq 0$ we break the SUSY of the problem thus removing the problem of the vanishing prefactor of the exponential contribution.
However the BM solution satisfies some SUSY Ward identities at any $\lambda_q$ and this ensures its physical consistency with respect to the problems discussed in the introduction. The correct way of treating the $\lambda_q=0$  case is by taking the limit $\lambda_q\rightarrow 0$, this is different from setting $\lambda_q=0$ from the beginning that instead yields the incorrect standard SUSY identities.
The SUSY Ward identities connect the bosonic order parameters of the theory with the fermionic ones. In this way they encode, within the standard computation of the TAP Complexity, the physical meaning, obtained through the cavity method, of the bosonic order parameters in terms of the physical observables $L$, $Z_1$, $Z_2$ that are indeed related  to averages of the macroscopic fermionic variables.

Although the results have been obtained within the SK model they can be extended to all models where a BM-like solution for the total complexity exists, {\it e.g.} the Ising $p$-spin model \cite{rieger}. In these models there is transition from a non-SUSY solution at high free energies 
to the SUSY solution \cite{MR} at low free energies \cite{isp}. 
Actually in all known FRSB and 1RSB models the complexity at the lower band edge, given by the Parisi solution, is  SUSY; this could be related to the fact that in some sense the relevant parameter for the non-SUSY complexity is  the self-overlap and not the free energy; instead at the lower band edge the free energy must be the relevant parameter since we want to recover thermodynamics.

Let us also notice that it would be interesting to use the information obtained on the behavior of the TAP solutions upon changes of the external parameters to gain further insight in the longstanding questions of chaos in temperature and magnetic field (see {\it e.g.} discussion in \cite{R1}). 
\\
{\bf Aknowledgements}. It is a pleasure to thank G. Parisi for interesting discussions and constant support. This work is dedicated to Elsa Fubini (1909-2003).

\end{document}